\documentclass[preprint,amsmath,superscriptaddress,showpacs,showkeys,amssymb,aps,prb]{revtex4-1}
\usepackage{graphicx,epsfig,multirow,dcolumn,bm}

\begin{document}
\title{Absence of first order magnetic transition, a curious case of Mn$_{3}$InC}

\author{E. T. Dias}
\affiliation{Department of Physics, Goa University, Taleigao Plateau, Goa 403206 India}
\author{A. Das}
\affiliation{Solid State Physics, Division, Bhabha Atomic Research Centre, Trombay, Mumbai 400085}
\author{A. Hoser}
\affiliation{Helmholtz-Zentrum Berlin, 14109, Berlin, Germany}
\author{S. Emura}
\affiliation{Institute of Scientific and Industrial Research, Osaka University, Osaka, Japan}
\author{A. K. Nigam}
\affiliation{Tata Institute of Fundamental Research, Dr. Homi Bhabha Road, Colaba, Mumbai 400005, India}
\author{K. R. Priolkar}
\affiliation{Department of Physics, Goa University, Taleigao Plateau, Goa 403206 India}
\email{krp@unigoa.ac.in}

\begin{abstract}
The volume expanding magnetostructural transition in Mn$_3$GaC and Mn$_3$SnC has been identified to be due to distortion of Mn$_6$C octahedra. Despite a similar lattice volume as Mn$_3$SnC and similar valence electron contribution to density of states as in Mn$_3$GaC, Mn$_3$InC does not undergo a first order magnetostructural transformation like the Ga and Sn antiperovskite counterparts. A systematic investigation of its structure and magnetic properties using probes like x-ray diffraction, magnetization measurements, neutron diffraction and extended x-ray absorption fine structure (EXAFS) reveal that though the octahedra are distorted resulting in long and short Mn -- Mn bonds and different magnetic moments on Mn atoms, the interaction between them remains ferromagnetic. This has been attributed to the strain on the Mn$_6$C octahedra produced due to relatively larger size of In atom compared to Sn and Ga. The size of In atom constricts the deformation of Mn$_6$C octahedra giving rise to Mn -- Mn distances that favor only ferromagnetic interactions in the compound.
\end{abstract}
\date{\today}

\maketitle

\section {Introduction}

In ternary manganese carbides and nitrides ordering with a cubic antiperovskite (Mn$_{3}$AB) crystal structure, the magnetic ground state and physical properties can be significantly influenced by merely replacing the A-site atom \cite{Bertaut19686,Antonov200775}. The A-site atoms like Ga, Cu, Zn, In or Sn form a cubic cage enclosing a Mn$_6$C octahedra at its center. Numerous reports have illustrated a fascinating array of exotic properties displayed by such Mn antiperovskites just by replacing the A-site atom. For example, a tunable negative or zero thermal expansion (NTE) accompanying the volume discontinuous magnetic transition found in most of the nitrides (Mn$_{3}$AN, A = Ga, Zn, etc.) is exceptionally absent in Mn$_{3}$CuN \cite{Takenaka200587,Matsuno200994,Sun201093}. A substitution of Cu by Ge atoms favorably alters the number of valence electrons to induce the NTE property in Mn$_{3}$Cu$_{0.5}$Ge$_{0.5}$N around room temperature \cite{Sun200942}. Even in carbides the first order magnetic transition seen at about 170 K in Mn$_3$GaC \cite{Bouchaud196637,Fruchart197844,Dias2014363}, disappears completely with replacement of Ga by Zn or Ge \cite{Fruchart197844}. The nature of magnetocaloric effect is also different in Mn$_3$GaC and Mn$_3$SnC. While Mn$_3$GaC displays an inverse magnetocaloric effect \cite{Cakir2012100,Dias2014363}, a normal magnetocaloric effect is seen in Mn$_3$SnC \cite{Wang200985,Dias201548}.

Recent extended x-ray absorption fine structure (EXAFS) studies on antiperovskites Mn$_{3}$GaC and Mn$_{3}$SnC suggest that the magnetic and magnetocaloric properties mainly originate from local distortions restricted to the Mn sublattice \cite{Dias2017122,Dias201548}. Even these distortions of the Mn$_6$C octahedra appear to be dependent on the type of A-site atom. In Mn$_3$GaC, the 8 fold degenerate Mn -- Mn bond split into long ($\sim$ 3.1 \AA) and short ($\sim$ 2.74 \AA) distances at $T_{C}$ = 242 K. The distortions are such that the Mn atoms are displaced from their face centered positions on a circular arc of radius equal to Mn -- C bond length. An abrupt decrease in the shorter Mn -- Mn distance at $T \sim$ 175 K results in an AFM ground state and gives rise to a large positive magnetic entropy change ($\Delta$S$_{M} \sim$ 15 J/kg-K at 2 T, equivalent to an adiabatic temperature difference of 3 K) \cite{Dias2014363,Dias2017122}. Introduction of Sn for Ga at the A-site results in a lattice expansion as well as a change in the nature of magnetocaloric effect from inverse to the conventional type with $\Delta{S}_{M}$ is about $-3$ J/kg-K at 2 T applied field \cite{Dias2015117}. The larger size of Sn also seems to affect the distortions so as to cause the Mn$_{6}$C octahedra to elongate along one direction and shrink along the other two. As a result the magnetic propagation vector, $k$ is $\frac{1}{2}, \frac{1}{2}, 0$ in Mn$_3$SnC as against $\frac{1}{2}, \frac{1}{2}, \frac{1}{2}$ in Mn$_{3}$GaC \cite{Dias201548}.

The differences in the distortions of Mn$_6$C octahedra are preserved even in the solid solutions of Mn$_3$Ga$_{1-x}$Sn$_x$C and result in lattice strain \cite{Dias2018124} as well as a cluster glassy ground state \cite{Dias201795}. Such a non ergodic ground state in Mn$_3$Ga$_{0.45}$Sn$_{0.55}$C was explained to be due to formation of Ga rich and Sn rich clusters in the compound. Formation of such clusters could be due to some unique property of the A-site atom. Ga and Sn have two distinct differences, firstly their size and second is their contribution of the valence electron density of states. In order to understand the exact cause of such distortions we chose to investigate the structural and magnetic properties of Mn$_3$InC. It has a similar lattice volume as Mn$_3$SnC while In contributes the same number of valence electrons as Ga. Furthermore band structure calculations indicated that Mn$_3$InC should have similar properties as Mn$_3$GaC \cite{Motizuki198821,Ishida199332}. Experimental reports on Mn$_3$InC are of conflicting nature. Early reports indicate Mn$_{3}$InC to undergo a first order transition to a complex magnetic state that is similar to that of its Sn counterpart below $T_{C} \sim$ 272 K \cite{Kanomata1991126,Kanomata199296}, while recent studies report a ferromagnetic ground state with a $T_C \sim$ 350 K followed by a transition to antiferromagnetic state at about 140 K \cite{Malik201532}.

Therefore, in order to understand the exact nature of ground state in Mn$_3$InC we report results of a systematic investigation on the structural and the magnetic properties of Mn$_3$InC. Further we use it as a prototypical compound to investigate the exact role of the A-site atom in the magnetostructural transformation seen in the Mn based antiperovskites. We show that the distortions on Mn$_6$C octahedra are dependent on the size of A-site atom. Comparatively, when the A-site is occupied by a smaller atom like Ga, the Mn$_6$C octahedra distort maximum resulting in a wider separation between long and short Mn -- Mn bonds and an antiferromagnetic ground state. Larger A-site atoms like In however, constrain the distortions of Mn$_6$C octahedra such as to prevent formation of shorter Mn -- Mn bonds that favor antiferromagnetic alignment and thus resulting in a ferromagnetic ground state.

\section {Experimental}
To synthesize polycrystalline Mn$_{3}$InC using the solid state reaction method, stoichiometric weights of powdered Mn and graphite were first thoroughly mixed with elemental In before the addition of 15\% excess graphite powder. The resulting mixture was then pressed into a pellet, encapsulated in an evacuated quartz tube and heated to 1073 K for the first 48 hr before being annealed at 1150 K for the next 120 hr \cite{Dias20141}. On cooling to room temperature,  the crystallographic symmetry and phase purity of the compound formed was identified with the help of an x-ray diffraction (XRD) pattern recorded using Mo $K_\alpha$ radiation. Next, the magnetic properties of the prepared antiperovskite were determined by various field (0.01 T and between $\pm$7 T) and temperature dependent (5 K to 500 K) measurements carried out in a Vibrating Sample Magnetometer (Quantum Design). Further, temperature dependent variations in the crystallographic and magnetic structures were traced by Rietveld analysis \cite{Carvajal1993192} of neutron powder diffraction patterns recorded on the PD-2 diffractometer ($\lambda$ = 1.2443 \AA) at Dhruva reactor, Bhabha Atomic Research Center, Mumbai, India and E6 powder diffractometer ($\lambda$ = 2.4 \AA) at the BER II reactor at Helmholtz Zentrum Berlin, Germany. In an attempt to comprehend the observed magnetic behavior of this material, a detailed examination of the local structure surrounding the Mn$_{6}$C octahedra is carried out by analyzing room temperature EXAFS spectra recorded in transmission mode at 9C and NW10A at Photon Factory Synchrotron Source, Tsukuba, Japan. During each measurement, both incident and transmitted intensities were simultaneously measured at the Mn/In K edges within the -200 to 1300 eV range using an ionization chamber filled with appropriate gases. Further, the EXAFS ($\chi$(k)) signal is extracted by reducing the K edge data using well established procedures in the Demeter program \cite{Ravel200512}.

\section {Results and discussion}
\begin{figure}[h]
\center
\includegraphics[width=\columnwidth]{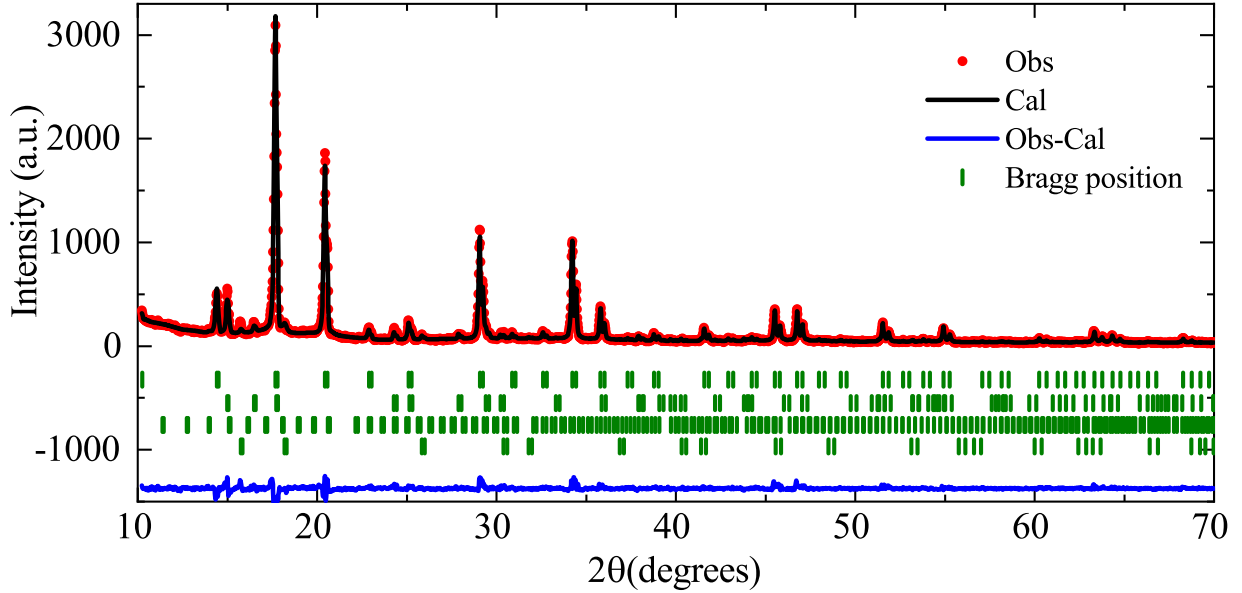}
\caption{Rietveld refined room temperature x-ray diffraction pattern recorded for Mn$_{3}$InC in the 10$ ^{\circ}$ $\leq 2\theta \leq$ 70$ ^{\circ}$ angular range.}
\label{fig:xrd}
\end{figure}

Rietveld analysis of the room temperature x-ray diffraction pattern recorded using Mo K$_\alpha$ ($\lambda$ = 0.7107 \AA~) radiation and presented in Fig. \ref{fig:xrd} indicates that the compound  Mn$_{3}$InC forms within the cubic antiperovskite structure (\emph{Space Group: Pm$\bar3$m}) along with impurities of In, In$_{2}$O$_{3}$ and MnO. The Mn to In ratio was found  to be 3:1 from the refinement of site occupancy of Mn and In in the antiperovskite phase.  This ratio along with C content \cite{Amos200273} was again independently verified from refinement  of neutron diffraction pattern described later in the paper. The structural and refinement parameters obtained for x-ray diffraction pattern are provided in Table \ref{tab:xray}.


\begin{figure}[h]
\center
\includegraphics[width=\columnwidth]{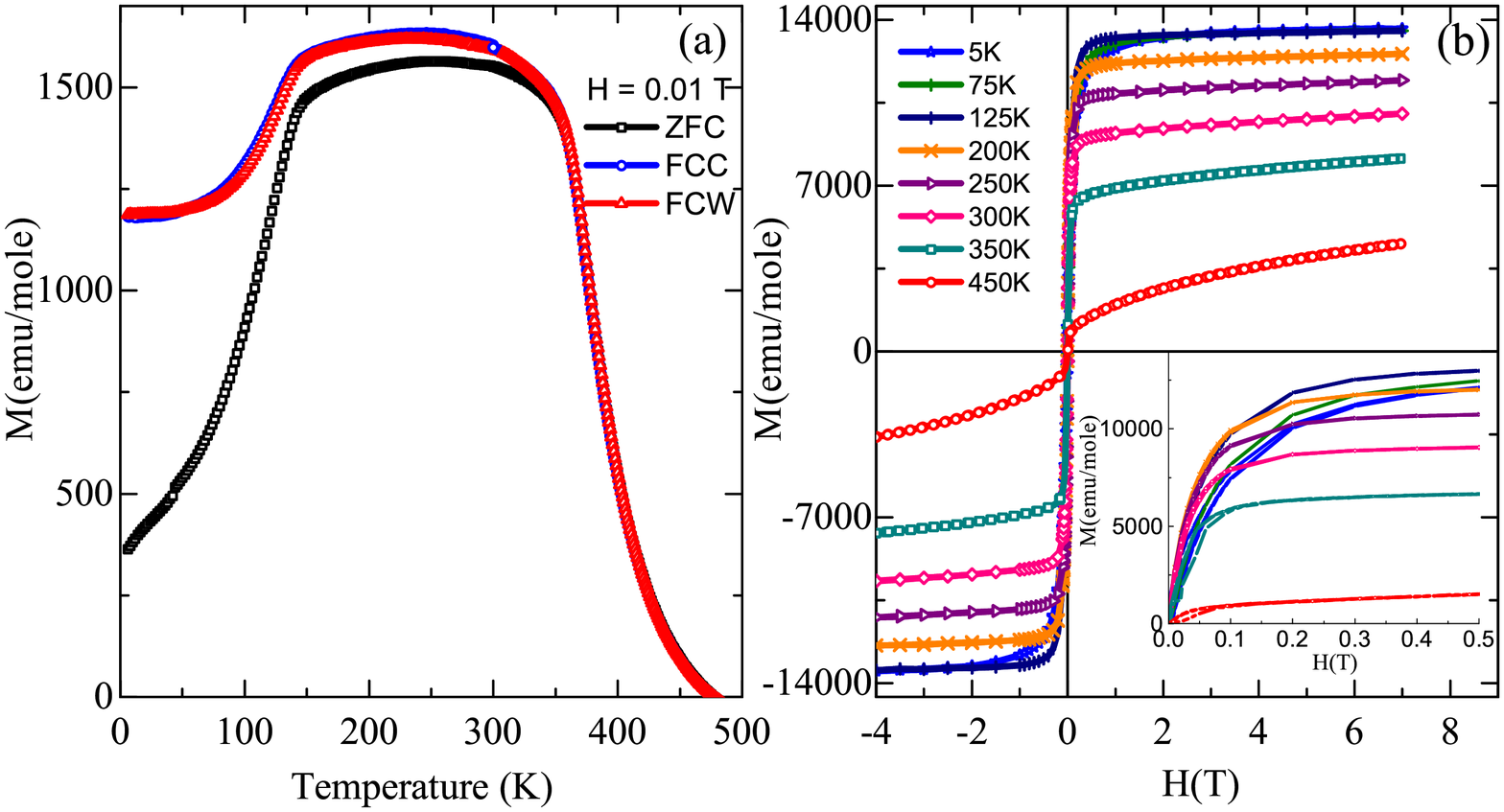}
\caption{(a) Magnetization data recorded as a function of temperature for Mn$_{3}$InC in the 5 -- 500 K temperature range under $H$ = 0.01 T. (b) Field dependent magnetization curves for Mn$_{3}$InC in the 5 K $\leq  T  \leq$ 450 K temperature range, $H$ = $\pm$7 T. Inset shows the behavior of virgin curve  at low fields}
\label{fig:mag}
\end{figure}

Temperature dependent magnetization curves (M(T)) presented for the compound in Fig. \ref{fig:mag} (a), over the 5 $-$ 500 K temperature range, in an applied field of 0.01 T, under both zero field cooled (ZFC) and field cooled (FCC and FCW) protocols exhibits a temperature dependence that differs from both Mn$_{3}$GaC and Mn$_{3}$SnC. The M(T) curves suggests that the compound exhibits a broad transition from a PM to a FM state at $T_{C} \sim$ 377 K followed by a sharp decrease in magnetization at T = 127 K resembling a transition to AFM state. Finite magnetic moment even under zero field cooled state has forced classification of this state to be consisting of both FM and AFM components \cite{Malik201532}. The FM character is also evident from the evolution of magnetization isotherms (M(H)) measured for the sample at several temperatures both above and below 377 K as well as 127 K. The M(H) curves recorded in magnetic fields up to $\pm$7 T in Fig. \ref{fig:mag} (b), exhibit a similar behavior resembling ferromagnetic order at all temperatures in the interval 5 K $\leq  T  \leq$ 350 K. An expanded view of the M(H) virgin curves at low fields shown in the inset of Fig. \ref{fig:mag} (b) indicates a weakening of ferromagnetic interactions below 100K. Further, presence of short range ferromagnetic interactions is also noted well above $T_C$ as indicated by the non linear M(H) curve at 450 K.

\begin{table}[htbp]
\caption{Crystal structure data and Rietveld refined parameters for Mn$_3$InC obtained from refinement of room temperature x-ray diffraction pattern recorded using Mo K$_\alpha$ radiation. Numbers on parenthesis indicate uncertainty in the last digit.}
\label{tab:xray}
\begin{tabular}{|c|c|c|c|}
\hline
\multicolumn{2}{|c|}{\textbf{Crystal system}} & \multicolumn{2}{|c|}{Cubic}\\ \hline
\multicolumn{2}{|c|}{\textbf{Space group}} & \multicolumn{2}{|c|}{Pm$\bar3$m}\\ \hline
\multicolumn{2}{|c|}{\textbf{a (\AA)}} & \multicolumn{2}{|c|}{3.99680(6)}\\ \hline
\textbf{Atom} & \textbf{Wyckoff} & \textbf{x, y, z} & \textbf{Occupancy}\\ \hline
Mn & 3c & 0.5, 0.5, 0 & 2.998(9)\\ \hline
In & 1a & 0, 0, 0 & 0.910(3)\\ \hline
C & 1b & 0.5, 0.5, 0.5 & 1 \\ \hline
\multicolumn{2}{|c|}{\textbf{Phase fraction  (\%)}} & \multicolumn{2}{c|}{\textbf{Reliability factors}}  \\ \hline
Mn$_{3}$InC & 92.58$\pm$0.72 & R$_{p}$ & 20.7  \\ \hline
In & 4.63$\pm$0.12 & R$_{wp}$ & 22.6    \\ \hline
MnO & 2.58$\pm$0.21 & R$_{exp}$ & 18.4   \\ \hline
In$_{2}$O$_{3}$ & 0.21$\pm$0.09 & $\chi^{2}$ & 1.266   \\ \hline
\end{tabular}
\end{table}

\begin{figure}[h]
\center
\includegraphics[width=\columnwidth]{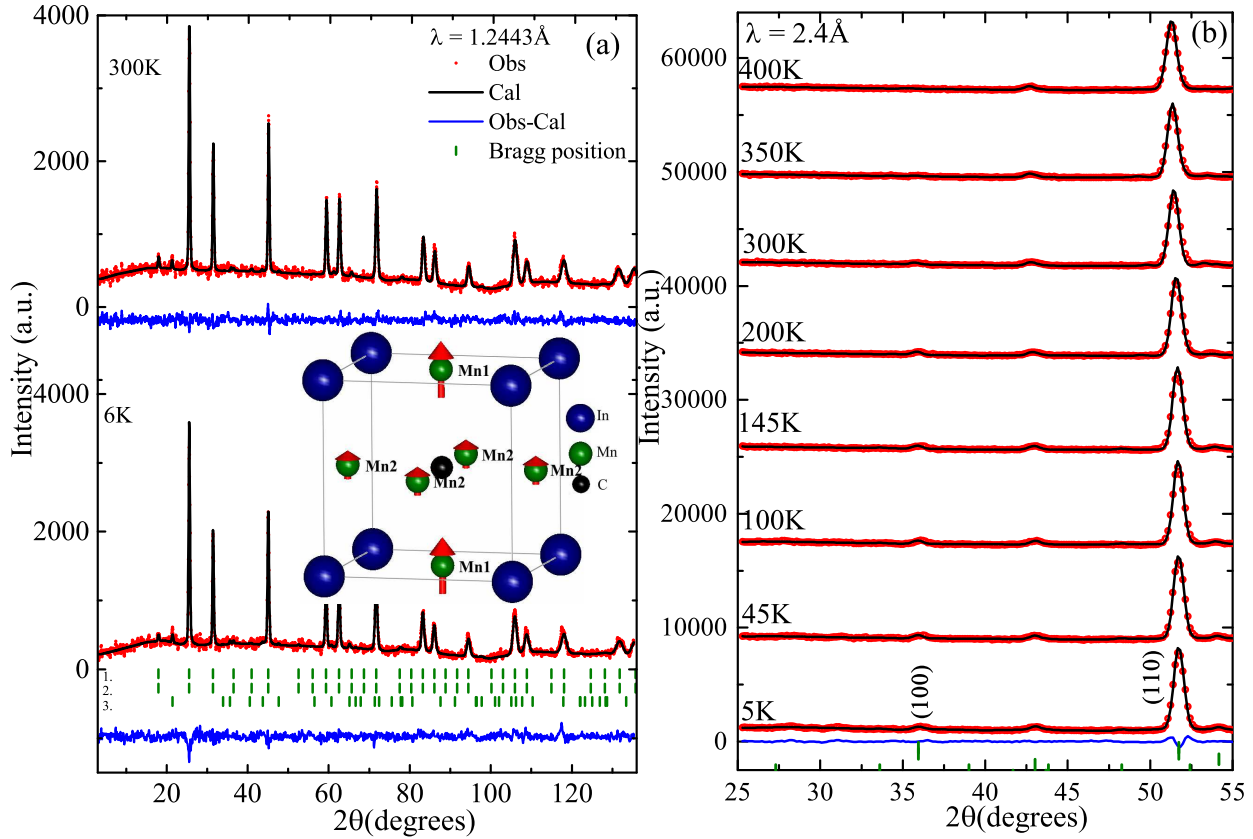}
\caption{(a) Temperature dependent neutron diffraction patterns recorded for Mn$_{3}$InC (with the Rietveld fit) in a neutron beam with $\lambda$ = 1.2443 \AA~. Bragg positions indicate crystallographic and magnetic phases of Mn$_{3}$InC along with excess graphite impurity. (b) Plot of neutron diffraction ($\lambda$ = 2.4 \AA~) patterns in the limited 2$\theta$ angular range at various temperatures between 5 K $-$ 400 K. Plots at higher temperatures have been artificially scaled up for clarity.}
\label{fig:nd1}
\end{figure}

In order to shed light on the exact nature of magnetic orderings in Mn$_{3}$InC, neutron diffraction patterns independently recorded in wavelengths $\lambda$ = 1.2443 \AA~ and $\lambda$ = 2.4 \AA~ were analyzed as a function of temperature. Rietveld refined structural parameters obtained from refining diffraction pattern recorded at 300 K using neutrons of 1.2443 \AA~ are given in Table \ref{tab:neutron}. The ratio of Mn:C obtained from neutron diffraction was found to be 3:1. Indium content was estimated slightly lower than that obtained from x-ray diffraction, but this could be due to larger absorption coefficient of In for thermal neutrons. It may also be seen that impurity phases of In and In$_2$O$_3$ are also not detected from neutron diffraction. Unlike its Ga and Sn counterparts that exhibit pure magnetic reflections below their respective transition temperatures \cite{Cakir2014115,Dias201548,Dias201795}, preliminary analysis of neutron diffraction data plotted in Fig. \ref{fig:nd1} (a) and (b) indicate no antiferromagnetic superlattice reflections at all temperatures down to 5 K. Instead, the presence of additional intensity in some of the low angle peaks confirms presence of long range ferromagnetic order in the compound. This additional intensity is present at all temperatures below 350 K, thus, ruling out presence of a second magnetic ordering transition at 127 K as indicated by magnetization measurements. Therefore, the decrease in ZFC magnetization at 127 K could be ascribed to a short range magnetic order or canting of Mn spins.

\begin{table}[htbp]
\caption{Structural and Rietveld parameters obtained from refinement of neutron diffraction pattern  of Mn$_3$InC recorded at 300 K using $\lambda$ = 1.2443 \AA~ radiation.}
\label{tab:neutron}
\begin{tabular}{|c|c|c|c|c|c|c|}
\hline
\textbf{Atom} & \textbf{Wyckoff} & \multicolumn{2}{c|}{\textbf{x, y, z}} & \multicolumn{2}{c|}{\textbf{Occupancy}}\\ \hline
Mn & 3c & \multicolumn{2}{c|}{0.5, 0.5, 0} & \multicolumn{2}{c|}{2.996(48)}\\ \hline
In & 1a & \multicolumn{2}{c|}{0, 0, 0} & \multicolumn{2}{c|}{0.791(13)}\\ \hline
C & 1b & \multicolumn{2}{c|}{0.5, 0.5, 0.5} & \multicolumn{2}{c|}{1.002(14)} \\ \hline
\multicolumn{6}{|c|}{\textbf{Magnetic structure}} \\ \hline
\multicolumn{3}{|c|}{\textbf{Magnetic k-vector}} & \multicolumn{3}{|c|}{(0, 0, 0)} \\ \hline
\textbf{Atom} & \textbf{x, y, z} & \textbf{m(a)} & \textbf{m(b)} & \textbf{m(c)} & \textbf{M(tot)}\\ \hline
Mn1 & 0.5, 0.5, 0 & 0 & 0 & 1.14 & 1.14$\pm$0.13 $\mu_B$\\ \hline
Mn2 & 0.5, 0, 0.5 & 0 & 0 & 0.57 & 0.57$\pm$0.18 $\mu_B$\\ \hline
Mn2 & 0, 0.5, 0.5 & 0 & 0 & 0.57 & 0.57$\pm$0.18 $\mu_B$\\ \hline
\multicolumn{3}{|c|}{\textbf{Phase fraction  (\%)}} & \multicolumn{3}{c|}{\textbf{Reliability factors}} \\ \hline
\multicolumn{1}{|c|}{Mn$_{3}$InC} & \multicolumn{2}{|c|}{ 99.82$\pm$1.41} & \multicolumn{1}{|c|}{R$_{p}$} & \multicolumn{2}{|c|}{ 25.7}\\ \hline
\multicolumn{1}{|c|}{C} & \multicolumn{2}{|c|}{ 0.18$\pm$0.02} & \multicolumn{1}{|c|}{R$_{wp}$} & \multicolumn{2}{|c|}{ 35.5}\\ \hline
\multicolumn{1}{|c|}{} & \multicolumn{2}{|c|}{} & \multicolumn{1}{|c|}{R$_{exp}$} & \multicolumn{2}{|c|}{14.2}\\ \hline
\multicolumn{1}{|c|}{} & \multicolumn{2}{|c|}{} & \multicolumn{1}{|c|}{$\chi^{2}$} & \multicolumn{2}{|c|}{3.258}\\ \hline
\end{tabular}
\end{table}


\begin{figure}[h]
\center
\includegraphics[width=\columnwidth]{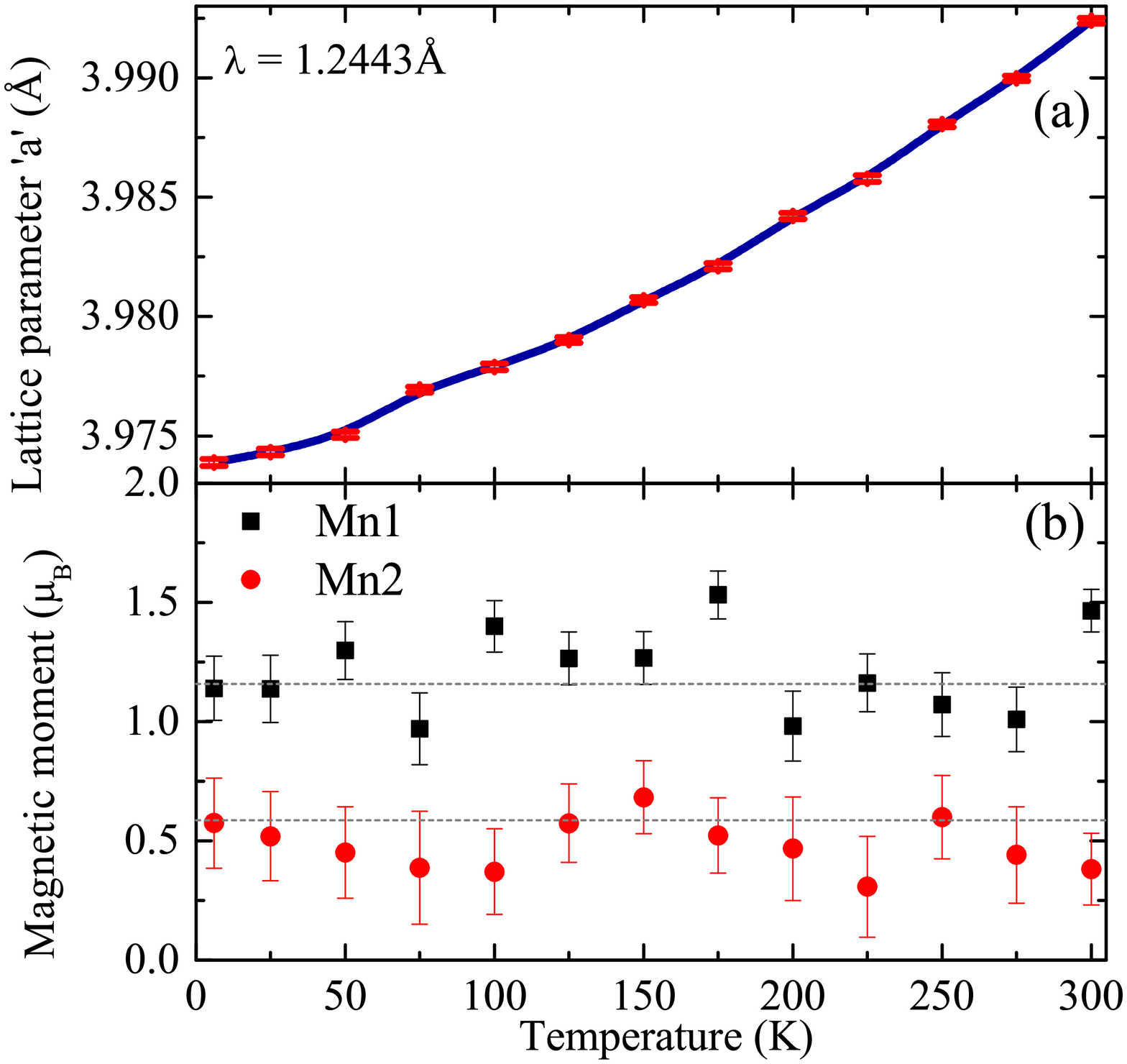}
\caption{(a) Variation of lattice parameter $a$ of Mn$_3$InC as a function of temperature. (b) Magnetic moment values of Mn1 and Mn2 as a function of temperature obtained from refinement of neutron diffraction patterns of Mn$_{3}$InC.}
\label{fig:nd2}
\end{figure}

In case of a FM structure the magnetic and nuclear reflections are observed at the same 2$\theta$ positions in neutron diffraction data. Consequently, the chemical and magnetic unit cells are identical. Using the basis vectors of the irreducible representations and appropriate magnetic symmetry operators generated by experimental propagation $k$ = $[0, 0, 0]$, the ferromagnetic structure obtained by trial and error method and its contributions to the umatched intensities were further identified. As shown in Fig. \ref{fig:nd1}, addition of a FM phase results in a satisfactory agreement between calculated and observed intensities of Rietveld refined diffraction patterns across the entire temperature range. The resulting magnetic structure consists of two FM sublattices identified as Mn1 and Mn2 with moment values of 1.14$\pm$0.13 $\mu_B$ and 0.57$\pm$0.18 $\mu_B$ respectively as shown in the inset of Fig. \ref{fig:nd1}. According to the temperature dependent variation of (100) and (110) Bragg reflections highlighted in the limited 2$\theta$ angular range in Fig. \ref{fig:nd1} (b) and the refined values of lattice parameter $a$ in Fig. \ref{fig:nd2}, the unit cell volume decreases monotonically with temperature and displays no discontinuity that can be associated to a first order transition. A slight change of slope observed in temperature variation of $a$ at $T \sim$ 127 K could possibly be associated with the broad magnetic transition seen in Fig. \ref{fig:mag} (a). Likewise, the thermal variation of magnetic moment calculated from neutron diffraction data for the two species of Mn atoms (see Fig. \ref{fig:nd2}(b)) also shows a smooth variation in the entire temperature range suggesting that the ferromagnetic ordering in Mn$_{3}$InC at $T_C$ = 377 K is of second order in nature. 

\begin{figure}[h]
\center
\includegraphics[width=\columnwidth]{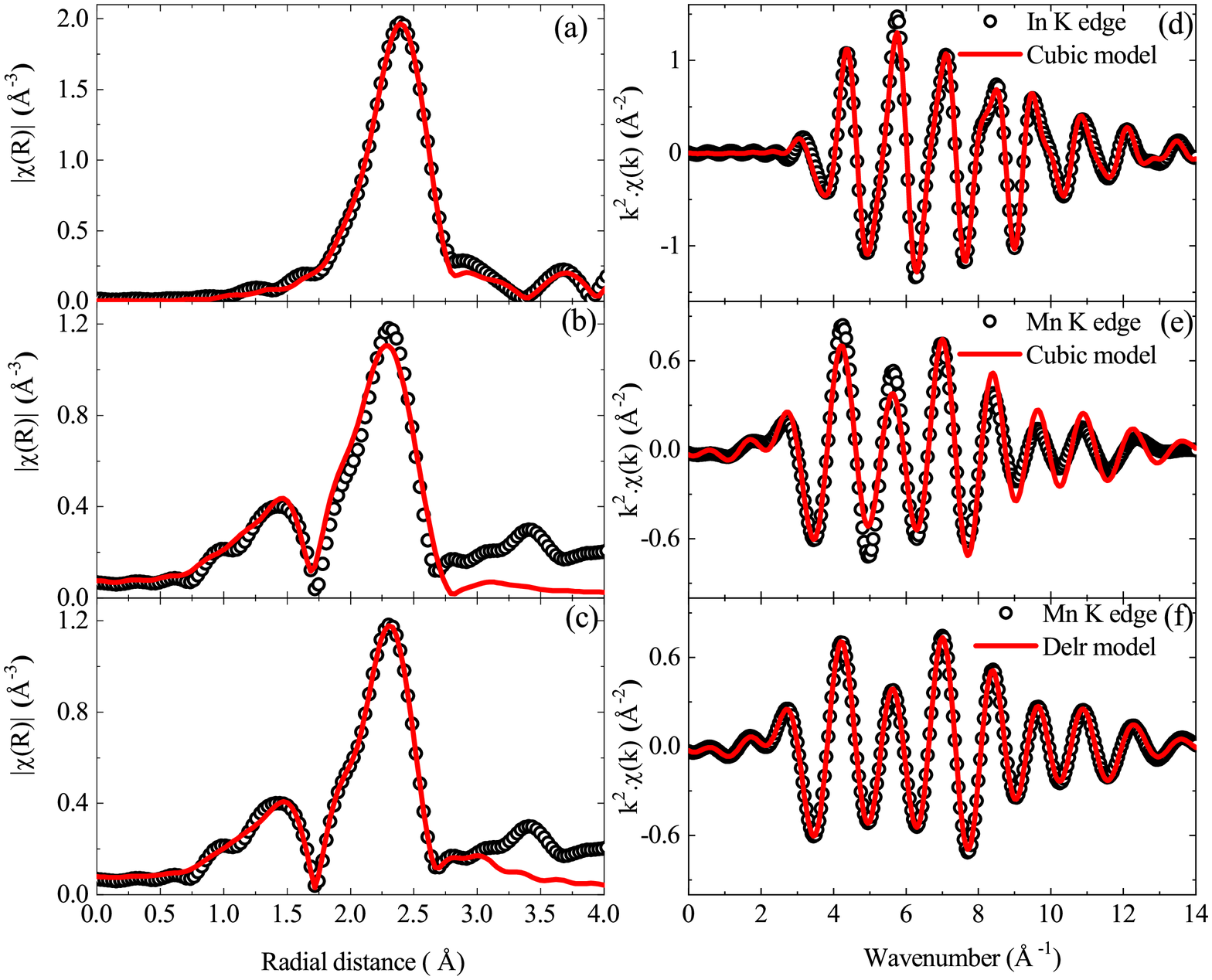}
\caption{Magnitude of the FT of $k^{2}$ weighted EXAFS spectra and the fitted curves at (a) In K edge and ((b) and (c)) at Mn K edge in Mn$_{3}$InC. The corresponding back-transformed spectra in $k$ space obtained from the $\chi(R)$ in the limited range of 1 to 3 \AA~ are shown for In edge (d) and Mn edge ((e) and (f)).}
\label{fig:xafs1}
\end{figure}

EXAFS studies contemplating the local structure of Mn$_{3}$Ga$_{1-x}$Sn$_{x}$C compounds have highlighted the strong relation between local strains introduced by the A-site atom on the Mn$_{6}$C octahedra and the magnetic properties exhibited by the respective compounds \cite{Dias201548,Dias201795,Dias2017122,Dias2018124}. While In atoms with an electronic configuration $[Kr]4d^{10}5s^{2}5p^{1}$, contribute the same number of valence electrons as Ga ($[Ar]3d^{10}4s^{2}4p^{1}$) to the Fermi level they have an average atomic size that is similar to Sn. Therefore, the non observation of a first order transition to AFM state in Mn$_{3}$InC raises fundamental questions on the structural distortions of Mn$_6$C octahedra introduced by In atoms at the A-site. Hence, in order to trace the changes in structural distortions in the local environment of the Mn$_{6}$C octahedra with the introduction In, room temperature EXAFS spectra were recorded and analyzed at the Mn and In K edges of Mn$_{3}$InC. Magnitudes of the $k^{2}$ weighted Fourier transformed (FT) EXAFS signals in the 3.0 $-$ 14.0 \AA~ $k$ range used for the analysis, are graphically represented in R = 0 to 4.0 \AA~ range in Fig. \ref{fig:xafs1}.

While the first peak in the Mn K edge data centered around R = 1.5 \AA~ is due to scattering from the nearest neighbor C atoms as in the case of Mn$_{3}$GaC and Mn$_{3}$SnC, the main peak in the R = 1.7 - 2.7 \AA~ range arises from the combined contribution from equidistant Mn and In next nearest neighboring atoms. To begin with, a cubic structural model described by crystal structure and structural parameters ($a$ = 3.9924 \AA~) obtained from room temperature neutron diffraction patterns, was used to fit the Mn K edge data in the 1 \AA~ to 3 \AA~ R range. While restrictions imposed by the cubic symmetry were forced upon the variation of the Mn -- C, Mn -- Mn and Mn -- In bond distances, the thermal mean square variation in bond distances ($\sigma^2$) were freely varied. The resulting fit in Fig. \ref{fig:xafs1} (b) highlights the clear deviation from cubic symmetry. Contradictorily, the same structural model used to analyze the In K edge EXAFS spectra in Fig. \ref{fig:xafs1} (a) (where the main peak in the range $R$ = 1.5 -- 3.0 \AA~ solely contains contributions from the In -- Mn correlation), results in a perfect fit. Thus, considering the similarities with results obtained for Mn$_{3}$GaC \cite{Dias2017122} and Mn$_{3}$SnC \cite{Dias201548}, where local distortions restricted to the Mn sublattice critically control the magnetic behavior exhibited by these materials, a structural model consisting of long and short Mn -- Mn distances was designed to fit the Mn K edge data of Mn$_{3}$InC. The resulting good fit in Fig. \ref{fig:xafs1} (c) implies presence of a structural distortion limited to Mn sublattice. The observed distortions are similar to those observed in Mn$_3$GaC wherein the Mn atoms are displaced from their face center positions on a circular arc of radius equal to Mn -- C bond distance and the values of long and short Mn -- Mn distances obtained respectively were 2.92 \AA~ and 2.76 \AA. Thus far the observations of XAFS study are quite similar to those observed in Mn$_3$GaC and Mn$_3$SnC and therefore still does not provide answer to non observation of first order transition as well as antiferromagnetic ordering in Mn$_3$InC.

\begin{figure}[h]
\center
\includegraphics[width=\columnwidth]{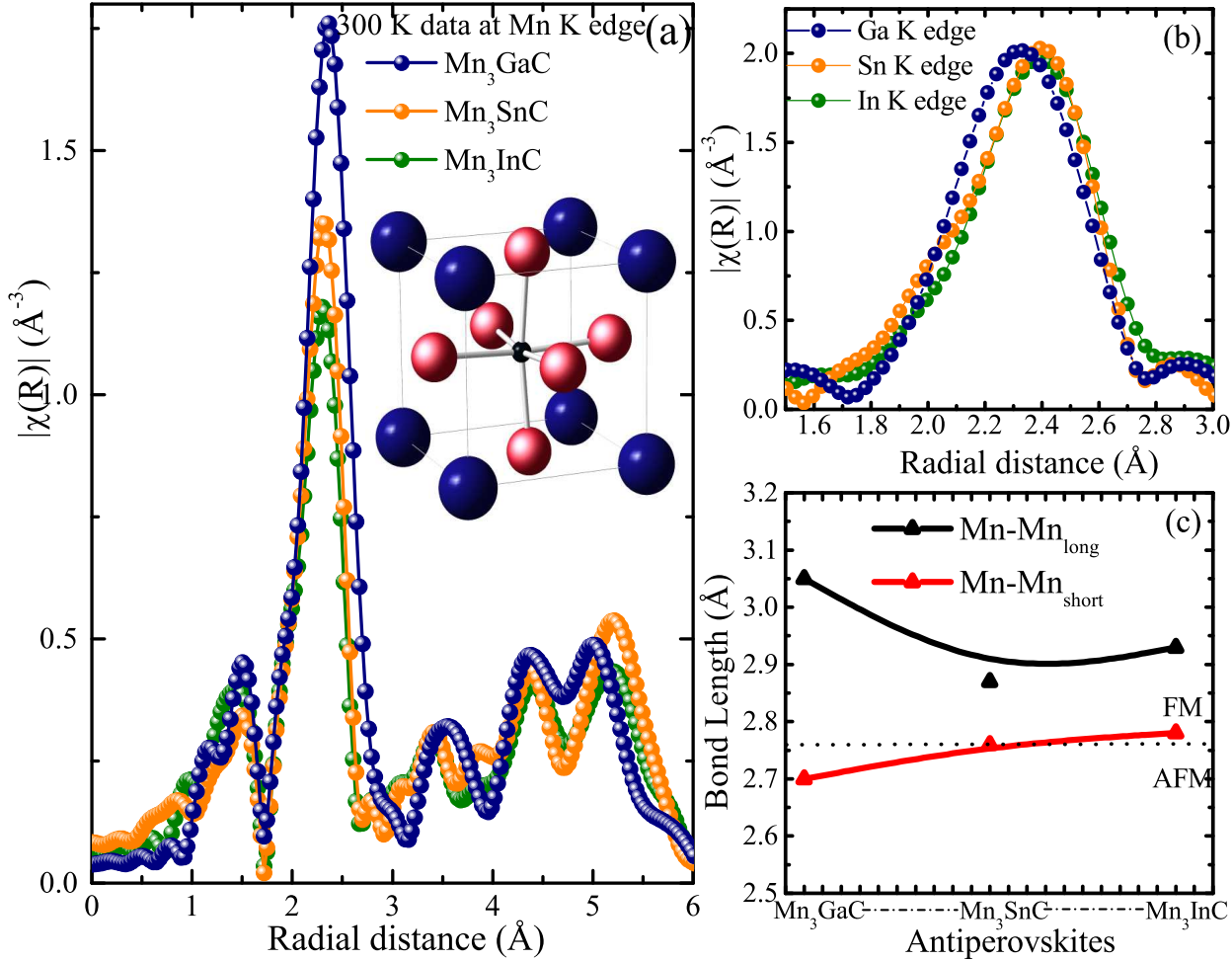}
\caption{(a) Variation of magnitude of $k^2$ weighted FT of EXAFS spectra at Mn K edge recorded at 300 K for Mn$_{3}$GaC, Mn$_{3}$SnC and Mn$_{3}$InC. Inset shows a schematic of local structural distortions. (b) Variation of magnitude of $k^2$ weighted FT of EXAFS spectra at the Ga, Sn and In K edges at 300 K. (c) Variation of Mn -- Mn bond distances in the three compounds.}
\label{fig:xafs2}
\end{figure}

A comparison between the Mn K EXAFS data recorded at 300 K in Mn$_{3}$GaC, Mn$_{3}$SnC and Mn$_{3}$InC in Fig. \ref{fig:xafs2} perfectly highlights the variation of local structural distortions brought about by the differing size of the A-site atom. The width of the main peak in the magnitude of FT of XAFS data at $\sim$ 2.3 \AA~ decreases from Mn$_3$GaC to Mn$_3$InC. The peak itself resembles a scattering correlation between the absorber atom and a scattering atom and the width of such a peak signifies presence of static (structural) as well as dynamic (temperature) disorder present in the compound. In the Mn K EXAFS, the peak at $\sim$ 2.3 \AA~ comprises of scattering from Mn -- Mn and Mn -- Ga, Sn or In correlations. While in Ga/Sn/In EXAFS the correlation at $\sim$ 2.3 \AA~ corresponds solely to Ga/Sn/In -- Mn scattering. Since the width of the first peak in Ga, Sn and In K EXAFS recorded at 300 K is nearly same (see Fig. \ref{fig:xafs2} (b)), any change in width of the peak in Mn K EXAFS is due to structural disorder in the Mn -- Mn correlations. It can be seen that the width of the peak in Mn K EXAFS decreases from that in Mn$_3$GaC to that in Mn$_3$InC implying thereby a decrease in local structural disorder.  When the A-site is occupied by a smaller atom like Ga, the Mn$_{6}$C octahedra distort freely resulting in a large separation between Mn -- Mn$_{long}$ ($\sim$ 3.1 \AA) and Mn -- Mn$_{short}$ ($\sim$ 2.7 \AA) bond distances. With the progressive increase in the size of A-site atom from Ga (r = 1.36 \AA) to Sn (r = 1.45 \AA) to In (r = 1.56 \AA), the distortions are constrained such that the difference between Mn -- Mn$_{short}$ and Mn -- Mn$_{long}$ decreases as illustrated in Fig. \ref{fig:xafs2}(c). This can be explained as follows: the size of In is about 14\% larger than that of Ga but the increase in lattice parameter is only about 3\%. This results in severe strain on the Mn$_{6}$C octahedra. Therefore despite presence of local structural distortion, the difference in Mn -- Mn$_{short}$ and Mn -- Mn$_{long}$ bond distances is much less. Also the both the Mn -- Mn distances being larger than 2.74 \AA, support only ferromagnetic interactions. However, the presence of distortions in Mn$_{6}$C octahedra leading to long and short Mn -- Mn distances provides an explanation to the observation of two magnetic sublattices in neutron diffraction analysis of Mn$_{3}$InC. Further the distortions of Mn$_{6}$C octahedra in Mn$_{3}$InC are more closer to the ones observed in Mn$_3$GaC then those seen in Mn$_{3}$SnC.  In Mn$_3$SnC the distortions are such that it results in longer and shorter Mn -- C bond distances \cite{Dias201548} which supports the presence of a small ferromagnetic moment on one of the Mn atoms and a normal magnetocaloric effect. In Mn$_{3}$InC such directional variations are not seen. Instead the Mn atoms are displaced on a circular arc of radius equal to Mn -- C distance. A similar but much larger distortions were seen in Mn$_{3}$GaC \cite{Dias2017122}. The similarities in octahedral distortions between Ga and In containing compounds could be due to similarities in their valence electron contribution to the band states near Fermi level.

\section{Conclusion}

In conclusion, systematic investigations on Mn$_{3}$InC reveal that it undergoes only a paramagnetic to ferromagnetic transition with a high $T_{C}$ = 377 K. EXAFS studies suggest that, due to the larger size of In atoms, the Mn$_{6}$C octahedra are strained. These strains restrict the distortions of the Mn sublattice such that the difference between Mn -- Mn long and short bond distances is smaller compared to that in Mn$_3$GaC and Mn$_3$SnC and the value of shorter Mn -- Mn distance is such that it does not favor antiferromagnetic ordering of the Mn sublattice. Thus it appears that the magnetic ground state in such antiperovskites is decided by the distortions of the Mn sublattice which are susceptible to the strain produced by the size of A-site atom.

\section*{Acknowledgment}
Council for Scientific and Industrial Research (CSIR), New Delhi is gratefully acknowledged for financial assistance under 03(1343)/16/EMR-II. Authors thank Photon Factory, KEK, Japan for beamtime on beamlines 9C and NW10A for the proposal No. 2014G042.

\bibliographystyle{apsrev4-1}
\bibliography{References}

\end{document}